\documentclass[aps,prl,reprint,superscriptaddress,floatfix]{revtex4-1}

\usepackage{amssymb}
\usepackage{amsmath}
\usepackage{psfrag}
\usepackage{topcapt}
\usepackage{hyperref}

\usepackage[utf8]{inputenc}

\newcommand{\bra}[1]{|#1\rangle}

\newcommand{\ghz}{~\mathrm{GHz}}
\newcommand{\mhz}{~\mathrm{MHz}}

\usepackage{ifpdf}

\ifpdf
\usepackage[pdftex]{graphicx}
\else
\usepackage{graphicx}
\fi

\begin{document}
\title{High-Fidelity Readout in Circuit Quantum Electrodynamics Using the Jaynes-Cummings Nonlinearity}

\author{M. D. Reed}
\affiliation{Departments of Physics and Applied Physics, Yale University, New Haven, Connecticut 06520, USA}

\author{L. DiCarlo}
\affiliation{Departments of Physics and Applied Physics, Yale University, New Haven, Connecticut 06520, USA}

\author{B. R. Johnson}
\affiliation{Departments of Physics and Applied Physics, Yale University, New Haven, Connecticut 06520, USA}

\author{L. Sun}
\affiliation{Departments of Physics and Applied Physics, Yale University, New Haven, Connecticut 06520, USA}

\author{D. I. Schuster}
\affiliation{Departments of Physics and Applied Physics, Yale University, New Haven, Connecticut 06520, USA}

\author{L. Frunzio}
\affiliation{Departments of Physics and Applied Physics, Yale University, New Haven, Connecticut 06520, USA}

\author{R. J. Schoelkopf}
\affiliation{Departments of Physics and Applied Physics, Yale University, New Haven, Connecticut 06520, USA}

\date{\today}

\ifpdf
\DeclareGraphicsExtensions{.pdf, .jpg, .tif}
\else
\DeclareGraphicsExtensions{.eps, .jpg}
\fi

\begin{abstract}
	We demonstrate a qubit readout scheme that exploits the Jaynes-Cummings nonlinearity of a superconducting cavity coupled to transmon qubits.  We find that in the strongly-driven dispersive regime of this system, there is the unexpected onset of a high-transmission ``bright'' state at a critical power which depends sensitively on the initial qubit state.  A simple and robust measurement protocol exploiting this effect achieves a single-shot fidelity of 87\% using a conventional sample design and experimental setup, and at least 61\% fidelity to joint correlations of three qubits.
\end{abstract}

\maketitle

Circuit quantum electrodynamics (cQED) is the study of the interaction of light and matter where superconducting qubits playing the role of atoms are strongly coupled to microwave transmission line resonators acting as cavities \cite{Wallraff2004}.  This architecture offers advantages over cavity QED with atomic systems with regard to both coupling strength and accessibility of strongly driven nonlinear regimes \cite{Fink2008, Schuster2008, Bishop2009}.  For example, the high-power response of a cavity with a qubit in resonance has recently been studied, demonstrating the appearance of additional photon number ($n$) resonances with characteristic $\sqrt{n}$ spacing, in excellent agreement with theory \cite{Bishop2009}.  In this Letter, we investigate the behavior of a strongly driven cQED system in the dispersive regime, where four strongly coupled transmon qubits \cite{Schreier2008} are far detuned from the cavity.  We find that for increasing drive strength, the cavity response initially becomes strongly nonlinear and  continuously shifts down in frequency.  At a critical power, it reaches its bare frequency and transmission rapidly rises to a ``bright state''.  Importantly, this critical power is strongly dependent on the initial state of the qubit, providing for a simple high-fidelity qubit readout mechanism.  This scheme is also applicable to the joint measurement of several qubits simultaneously.

\begin{figure*}
\centering
\includegraphics[scale=1]{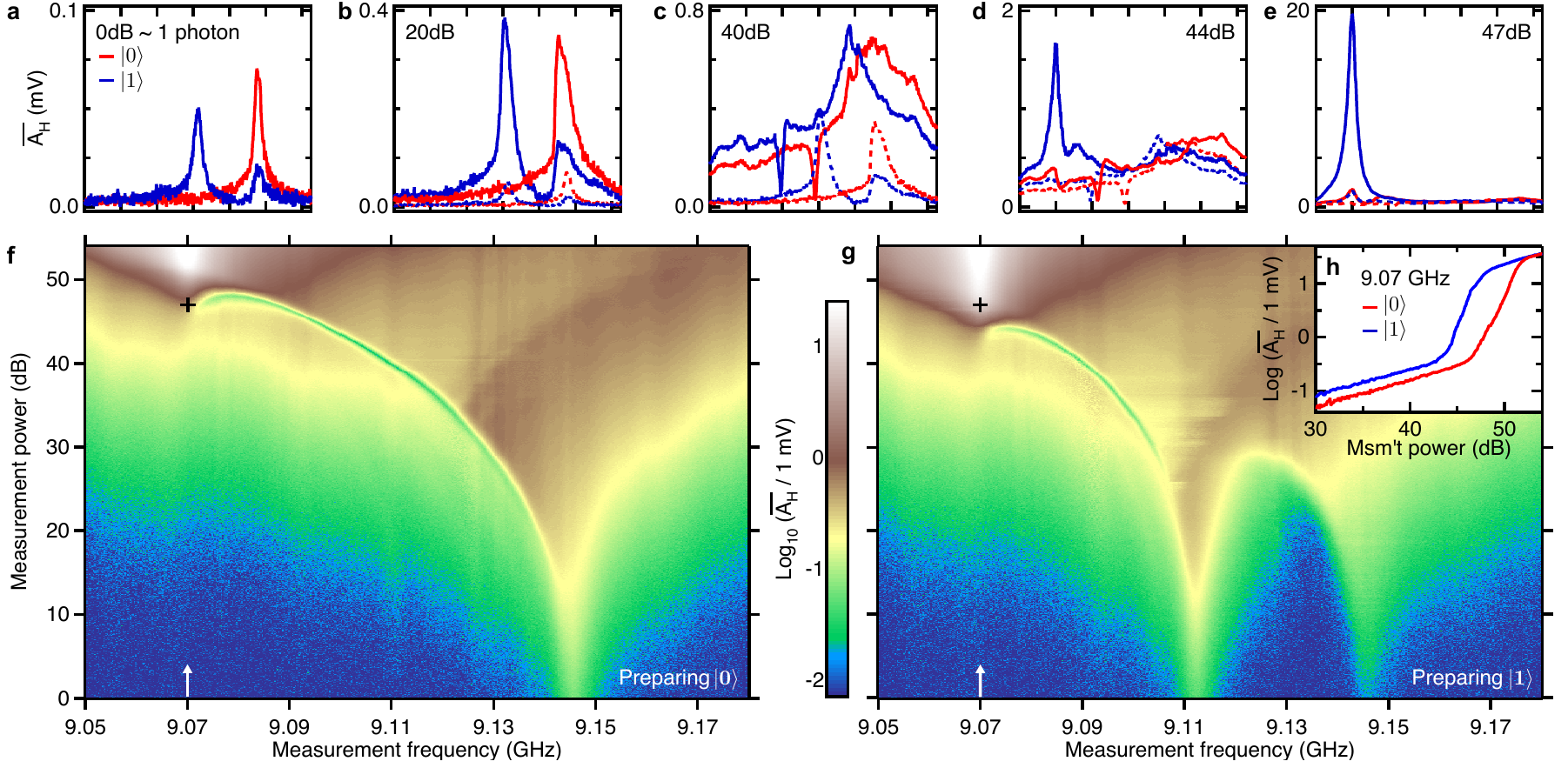}
\caption{
	(Color)
	Cavity transmission as a function of drive power and frequency, demonstrating the ``bright'' state at high incident power. 
	(a) Dispersively shifted cavity response for excited (blue) and ground (red) states of the $8\ghz$ qubit with $\sim1$~photon mean cavity occupation.  We reference this power to 0~dB.  For all plots in this figure, the $8\ghz$ qubit is prepared, a measurement tone is pulsed on, and the responding homodyne amplitude is averaged for 400~ns to yield $\overline{A_{\mathrm{H}}}$.  The mV scale used is arbitrary, but consistent to ease comparison.  The $x$-axis for (a-e) is frequency and covers the same range as for (f-g).
	(b-d) Cavity response for increasing drive power, with data for previous power plotted with dashed lines.  Transmission is inhibited by the cavity's inherited nonlinearity, limiting dispersive measurement fidelity.  Note for example that increasing drive 10~dB from (b) to (c) increases $\overline{A_{\mathrm{H}}}$ by only a factor of $\sim2$ and complicates the frequency dependence.  The emergence of a distinct resonance can be seen in (d).
	(e) High-transmission bright state.  At large drive power, the Jaynes-Cummings cavity anharmonicity shrinks sufficiently to allow near-unity transmission at $f_{\mathrm{bare}}$.  For this power the system only reaches its bright state when the qubit is excited due to the asymmetry of the dispersive cavity shift about $f_{\mathrm{bare}}$.  This asymmetry is characteristic to the transmon qubit \cite{Schreier2008} but might be possible to simulate for other designs \cite{Bishop2010}.
 	(f-g) Cavity response (log magnitude) for qubit ground and excited states.  The cavity continuously evolves from its low power linear behavior through the anharmonic bistable region and to the bright state.  There are two peaks present in (g) due to qubit relaxation during measurement.  The symbols (+) denote the optimal power and frequency for qubit readout, where the cavity response for the two qubit states is maximally different.
	(h) Response at $f_{\mathrm{bare}}$ (arrows) versus input power, showing a steep jump in transmission corresponding to the onset of the bright state at a qubit-state-dependent power.  Though transmission state dependence exists elsewhere, the behavior here is especially amenable for use as a qubit readout because the difference is large compared to amplifier noise.
	}
{\label{fig:pspec1q}}
\end{figure*}

There are two commonly used cQED readout schemes.  The first employs the state-dependent dispersive shift of the coupled cavity's resonance frequency [Fig.~\ref{fig:pspec1q}(a)] to infer the qubit state by measuring transmission \cite{Blais2004,Wallraff2005}.  This must be done in the low-power, linear response regime of the cavity in part because system anharmonicity inhibits higher population.  The signal-to-noise ratio (SNR) and fidelity of this weak measurement is thus limited (to typically 40\%-60\% \cite{Wallraff2005, Johnson2010}) by the relatively high noise temperature ($T_{\mathrm{N}}^{\mathrm{sys}}\approx10 ~\mathrm{K}\sim 20~\mathrm{photons}$) of conventional commercial amplifiers and short integration time mandated by qubit relaxation ($T_1 \approx 1~\mu s $) \cite{Gambetta2007}.  A specialized superconducting low-noise amplifier \cite{Spietz2009, Castellanos2009, Bergeal2009} could improve the SNR, but at a cost of additional complexity.  The second scheme uses some nonlinearity to project the qubit state onto a classically distinguishable system and yield high discrimination \cite{Siddiqi2006}.  The Josephson bifurcation amplifier (JBA) is one such example, exploiting cavity bistability caused by the Kerr-Duffing nonlinearity of an additional Josephson junction \cite{Siddiqi2004, Boulant2007, Mallet2009}.  These schemes have, however, not yet taken advantage of the nonlinearity inherent to the dispersive interaction.  The mechanism we report here also projects the qubit state onto a classical state of the system but does not require additional Josephson junctions or any other hardware modifications.  It instead employs the nonlinearity due to the Jaynes-Cummings interaction of the qubit and cavity, in qualitative agreement with theory \cite{Bishop2010, Boissonneault2010}, effectively using the qubit as its own amplifier.

Our readout scheme exploits the unusual behavior of the system when driven strongly.  For low drive strength, the cavity (at $f_{\mathrm{bare}} = 9.070\ghz$) is dispersively shifted many linewidths by the strongly coupled transmons at frequencies $(f_1, f_2, f_3, f_4) = (6.000, 7.000, 8.000, 12.271)\ghz~\pm~2\mhz$ (see Supplement for device details).  The cavity also inherits anharmonicity ($\alpha$, the frequency difference between subsequent cavity transitions) from the interaction, which inhibits transmission for larger power [Figs.~\ref{fig:pspec1q}(a-d)].  For small number of excitations $n$, $\alpha$ is to first order constant \cite{Milburn1995}, though higher order terms are often considered as $n$ approaches a critical value \cite{Boissonneault2009}.  This expansion neglects the high-power behavior of the system, which can be recovered with a semiclassical model, shown previously for the case of a qubit and cavity in resonance \cite{Peano2010}.  In a parallel publication \cite{Bishop2010}, Bishop {\it et al.} consider the case of a large cavity-qubit detuning semiclassically and show that for increasing $n$, $\alpha$ decreases monotonically and the cavity frequency ($f_c$) shifts toward $f_{\mathrm{bare}}$, in qualitative agreement with our observations.  The measured cavity output power as a function of incident drive power and frequency thus smoothly evolves from a linear relation at low occupation ($n\sim1$), where the shift due to increasing $n$ is small compared to cavity linewidth, to a nonlinear relation at intermediate $n$ where $\alpha$ is important [Fig.~\ref{fig:pspec1q}(f)].  A dip in transmission likely caused by interference between two bistable solutions follows the cavity down in frequency with increasing power and disappears at $f_{\mathrm{bare}}$.  At large drive power, the cavity reaches a near-unity transmission ``bright state'' where $f_c \to f_{\mathrm{bare}}$ and $\alpha \to 0$.  This power is $\sim 50,000$ times larger than what induces a one-photon mean population, producing a signal large compared to measurement noise.  We reference the one-photon power to 0~dB.

\begin{figure}
\centering
\includegraphics[scale=1]{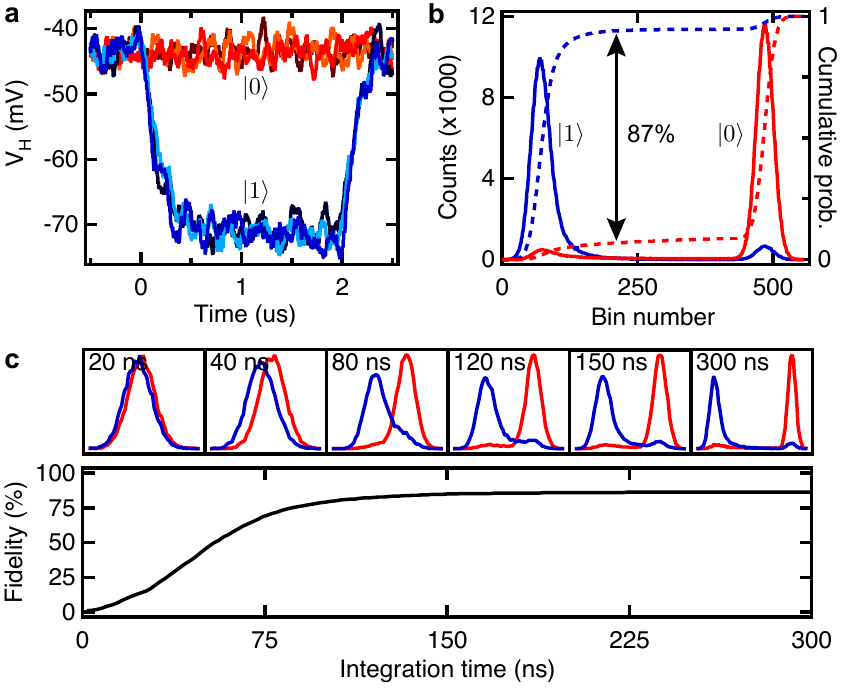}
\caption{
	Measurement transients and histograms for single-shot measurement of one qubit (at $8\ghz$).  
	(a) Three measurements of single-shot system response $V_{\mathrm{H}}(t)$ to driving at $f_{\mathrm{bare}}$ with a power $P_{\mathrm{meas}}$ for prepared qubit ground and excited states, low-passed to the response time of the cavity $1/\kappa\approx 100$~ns.  They are reproducible and well-distinguished, demonstrating that the response is large compared to the measurement noise.  
	(b) Histograms and S-curves quantifying measurement fidelity.  An ensemble of single-shot responses are integrated for 500~ns and their distribution plotted. The two histogram peaks (solid lines) are well separated with few counts between them.  Integrating these yields ``S-curves'' (dashed lines), with their maximal difference indicating a single-shot fidelity of 87\%.
	(c) Measurement fidelity versus integration time.  Integrating $V_{\mathrm{H}}(t)$ for 120~ns yields 80\% fidelity, while 240~ns yields 87\%.  The histogram evolution shows that integrating longer further separates the peaks but does not increase fidelity.
	}
{\label{fig:hist1q}}
\end{figure}

This behavior can be used to realize a high-fidelity qubit measurement because of a qubit-state-dependent bright-state onset power.  When driving with sufficiently high power at $f_{\mathrm{bare}}$, the system will excite to its bright state and remain there as long as the drive is applied.  Crucially, however, the lowest power for which the system will excite depends strongly on the qubit state because the cavity is dispersively shifted closer to $f_{\mathrm{bare}}$ when the qubit is excited [Fig.~\ref{fig:pspec1q}(g)].  This smaller detuning reduces the Lorentzian filtering effect of the cavity at $f_{\mathrm{bare}}$, lowering the onset power.  The cavity response at $f_{\mathrm{bare}}$ thus undergoes a sharp step for increasing drive at a power depending on the qubit state [Fig.~\ref{fig:pspec1q}(h)].  This transition does not happen discontinuously; there are powers where the system will only excite some of the time.  A high-discrimination readout is then made by choosing a drive power $P_{\mathrm{meas}}$ that maximizes the difference in bright-state probability for the qubit ground and excited states, generating a large state-dependent transmission difference [Fig.~\ref{fig:hist1q}(a)].  For measurements of the $8\ghz$ qubit, the optimal power is $P_{\mathrm{meas}} \approx +47~\mathrm{dB}$.  

We determine the readout measurement fidelity by comparing the ensemble distribution of cavity responses having prepared the qubit ground and excited states.  This is done by integrating $\sim500,000$ (in this case) individual measurement transients such as those shown in Fig.~\ref{fig:hist1q}(a) for $500~\mathrm{ns}$ for both qubit preparations and plotting their distribution [Fig.~\ref{fig:hist1q}(b)].  For the qubit at $8\ghz$, the maximal separation between the normalized integral of these histograms indicates a measurement fidelity of 87\%.  This  increases to 91\% when pulsing to the second transmon excited state just before measurement (data not shown; similar to Ref.~\onlinecite{Mallet2009}).  Fidelity is likely limited by qubit relaxation before its state is projected (false negatives) and spurious system excitations due to insufficiently separated onset powers (false positives).  Quantitative predictions of fidelity based on parameters like qubit lifetime are unavailable because the fast dynamics of the measurement process are not yet well understood.

Two important considerations for qubit readout are measurement time and repetition rate.  Single-shot fidelity as a function of integration time is shown in Fig.~\ref{fig:hist1q}(c).  After 120~ns the fidelity has risen to 80\%, and by 240~ns, it reaches its ultimate value of 87\%.  This time is due to the speed with which the system excites to the bright state and the SNR.  Integrating longer further separates the histogram distributions but does not change the fidelity because it is limited by failed encoding of the qubit state onto the cavity, not noise.  The large number of measurement photons used might be of concern regarding the creation of quasiparticles in or heating of the sample.  We have not observed any effect on qubit coherence when using this scheme, however, suggesting that any effects are gone after the 10~$\mu \mathrm{s}$ repetition time, nor does its use affect the base temperature of our $^3\mathrm{He}$-$^4\mathrm{He}$ dilution refrigerator (25~mK).  An example measurement of a Rabi oscillation is shown in Supplementary Fig. S1.

\begin{figure}
\centering
\includegraphics[scale=1]{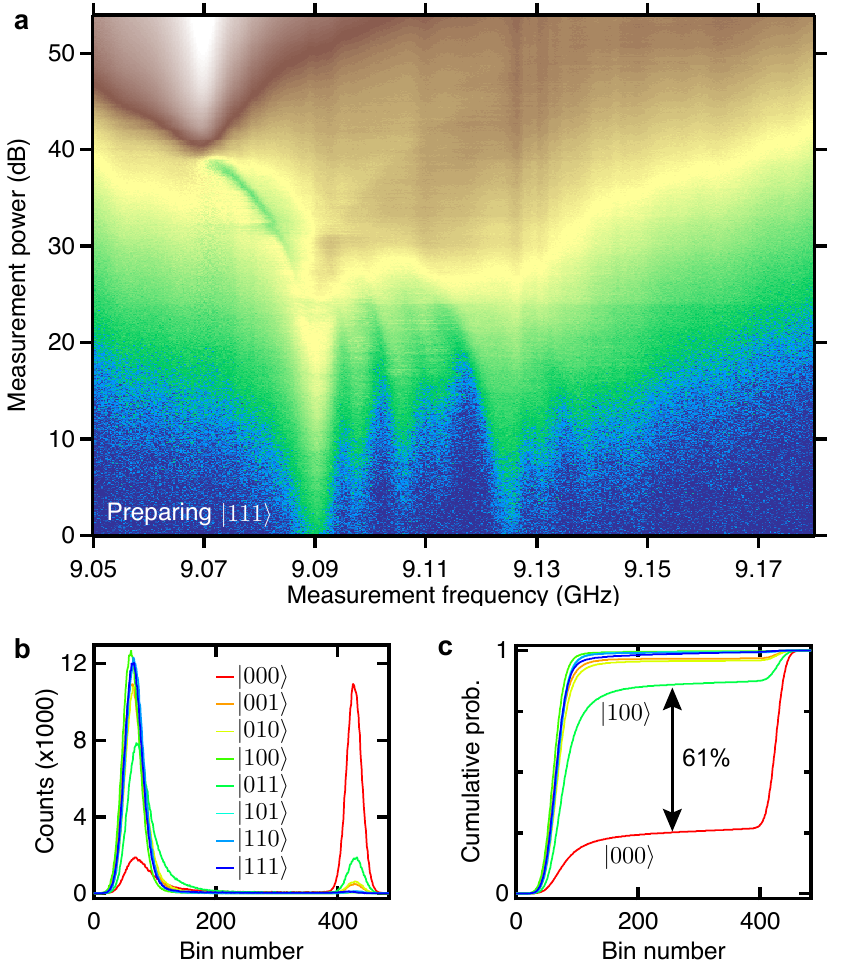}
\caption{ 
	Joint qubit readout.
	(a) Pulsed cavity response $\overline{A_{\mathrm{H}}}$ when preparing the $\bra{111}$ state.  The eight ($2^3$) register states of three qubits induce a different dispersive cavity shift, each discernible at low power due to decay of the $\bra{111}$ state during measurement.  These frequencies were independently measured to be $( f_{\bra{000}}, f_{\bra{100}}, f_{\bra{010}}, f_{\bra{001}}, f_{\bra{110}}, f_{\bra{101}}, f_{\bra{011}}, f_{\bra{111}} ) = ( 9.145, 9.139, 9.131, 9.112, 9.124, 9.105, 9.097, 9.090 )~\mathrm{GHz}$, with $f_{\mathrm{bare}}=9.070\ghz$.  The most prominent secondary cavity position corresponds to the third qubit (at $8\ghz$) relaxing during measurement, consistent with its Purcell-limited \cite{Houck2008} $T_{\mathrm{1}}$ being shortest of the three, with $T_{\mathrm{1}}^{\mathrm{Q}_{\mathrm{1}}}= 1.2 ~\mu \mathrm{s}$, $T_{\mathrm{1}}^{\mathrm{Q}_{\mathrm{2}}}= 1.0 ~\mu \mathrm{s}$, and $T_{\mathrm{1}}^{\mathrm{Q}_{\mathrm{3}}}= 0.6 ~\mu \mathrm{s}$.  The system excites to its bright state at lower power here than seen in Fig.~\ref{fig:pspec1q} because of the smaller initial cavity detuning from $f_{\mathrm{bare}}$.  The color scale is identical to that in Fig.~\ref{fig:pspec1q}.  
	(b) Histograms for all eight basis states, when driving with +49~dB at $f_{\mathrm{bare}}$.  
	(c) S-curves for joint qubit readout, indicating a minimum of 61\% fidelity for distinguishing the $\bra{000}$ from any other state.  The best fidelity is in this case obtained with a $\sim25\%$ probability of exciting the system when prepared in $\bra{000}$.
	}
{\label{fig:pspec3q}}
\end{figure}

We have extended this readout scheme to the joint readout of three qubits.  The physics of multiqubit readout are similar to the solitary case, but generalized for the eight zero-power cavity frequencies possible with three qubits.  Each basis state induces a different dispersive cavity shift from $f_{\mathrm{bare}}$, yielding a hierarchy of powers for the onset of the bright state.  Readout is done by choosing a measurement power above the onset power for all but the state one is trying to detect.  We consider the register of three qubits at frequencies $(6,7,8)\ghz$, with the fourth qubit unused.  For these detunings, $P_{\mathrm{joint}} \approx 49~\mathrm{dB}$ is optimal for distinguishing $\bra{000}$ from all other basis states (including second excited states).  This scheme is especially convenient for performing qubit tomography, which involves measuring multiqubit correlations \cite{Steffen2006} and is done efficiently with a joint qubit measurement \cite{DiCarlo2009, Filipp2009, Chow2010}.  Indeed, the creation and detection of tripartite entanglement was recently demonstrated using this device and readout scheme \cite{DiCarlo2010}.  Cavity transmission for the prepared excited states of all three qubits is shown in Fig.~\ref{fig:pspec3q}(a).

The joint fidelity is quantified in a similar manner as for one qubit, but expanded for the eight qubit basis states [Fig.~\ref{fig:pspec3q}(b)].  As there is no established definition of joint qubit readout fidelity, we propose a conservative metric: the maximal separation between the S-curves for the state one is trying to detect and the least distinguishable state.  This standard yields a fidelity of 61\% to the $\bra{000}$ state for these qubit detunings [Fig.~\ref{fig:pspec3q}(c)], limited by the distinguishability of the excited state of the first qubit (at $6\ghz$).  Fidelity could be substantially improved by either reducing that qubit detuning (at a cost of increased residual qubit-qubit coupling) or by pulsing to the second transmon excited state.  When an additional $\pi$ pulse between the first and second excited states is included, the joint fidelity is increased to $\sim80\%$ (data not shown).  This scheme could presumably be scaled to a larger number of qubits, subject to concerns of spectral crowding and residual qubit coupling.

This measurement scheme, which we name the Jaynes-Cummings readout, provides high-fidelity single-shot qubit measurement in cQED while being simple, relatively insensitive to qubit detuning, and not requiring any change in experimental setup or sample design.  It works by combining the occupation-dependent Jaynes-Cummings anharmonicity \cite{Bishop2010, Boissonneault2010, Carbonaro1979} with the qubit-state-dependent dispersive cavity shift to conditionally drive the system to a high-transmission bright state, in analogy to the fluorescence readout scheme used in ion traps \cite{Burrell2010}.  The effect has also delivered similar performance in a Purcell-filtered \cite{Reed2010} low-Q cavity coupled to a single transmon, suggesting that it is also robust to cavity lifetime.  It may also be extendable to other cQED architectures, not just those using transmons \cite{Bishop2010}.  We do not expect it to be quantum nondemolition due to dressed dephasing \cite{Boissonneault2008} of the qubit from the substantial photon population, but this is not necessary for many applications.  Future work will further optimize the readout scheme by changing sample design parameters.  A scheme may be possible where all qubit states are measured in a single shot by sweeping the measurement tone power and frequency or by using an intermediate power where the transition rate for each basis state is substantially different.

We thank L.~S.~Bishop, E.~Ginossar, M.~H.~Devoret, and S.~M.~Girvin for discussions and M.~Brink for experimental contributions.  We acknowledge support from LPS/NSA under ARO Contract No.\ W911NF-05-1-0365, from IARPA under ARO Contract No.\ W911NF-09-1-0369, and from the NSF under Grants No.\ DMR-0653377 and No.\ DMR-0603369. Additional support provided by CNR-Istituto di Cibernetica, Pozzuoli, Italy (LF).  All statements of fact, opinion or conclusions contained herein are those of the authors and should not be construed as representing the official views or policies of the U.S. Government.

\bibliographystyle{apsrev4-1}
\bibliography{readout_etal}

\end{document}


\title{Supplementary Material for ``High-Fidelity Readout in Circuit Quantum Electrodynamics Using the Jaynes-Cummings Nonlinearity''}

\author{M. D. Reed}
\affiliation{Departments of Physics and Applied Physics, Yale University, New Haven, Connecticut 06520, USA}

\author{L. DiCarlo}
\affiliation{Departments of Physics and Applied Physics, Yale University, New Haven, Connecticut 06520, USA}

\author{B. R. Johnson}
\affiliation{Departments of Physics and Applied Physics, Yale University, New Haven, Connecticut 06520, USA}

\author{L. Sun}
\affiliation{Departments of Physics and Applied Physics, Yale University, New Haven, Connecticut 06520, USA}

\author{D. I. Schuster}
\affiliation{Departments of Physics and Applied Physics, Yale University, New Haven, Connecticut 06520, USA}

\author{L. Frunzio}
\affiliation{Departments of Physics and Applied Physics, Yale University, New Haven, Connecticut 06520, USA}

\author{R. J. Schoelkopf}
\affiliation{Departments of Physics and Applied Physics, Yale University, New Haven, Connecticut 06520, USA}

\date{\today}

\ifpdf
\DeclareGraphicsExtensions{.pdf, .jpg, .tif}
\else
\DeclareGraphicsExtensions{.eps, .jpg}
\fi
\maketitle

\subsection{Device Fabrication and Parameters}

The device is fabricated on a C-plane sapphire ($\mathrm{Al_{2}O_{3}}$) substrate.  Optical lithography followed by fluorine-based reactive ion etching is used to pattern the coplanar waveguide (CPW) structure comprising the resonator onto sputtered niobium.  The resonance frequency of this structure, in the absence of the qubit dispersive shift, is the so-called ``bare'' cavity frequency $f_{\mathrm{bare}} = 9.070\ghz$.  Gaps in the CPW at either end of the chip comprise coupling capacitors to the environment, giving the cavity a linewidth $\kappa=2.4 \mhz$.  

Electron-beam lithography is used to pattern the four transmon qubits, which are deposited using a double-angle aluminum electron-beam evaporation process.  Maximum Josephson energies for qubits $Q_1$ to $Q_4$ are, respectively, $E^{\mathrm{max}}_{J}/h = (42, 29, 47, 57) \ghz$, with qubit-cavity coupling $g/2\pi\approx 220\mhz$ (not including the multiplicative level-dependent coupling matrix element; see the Methods section of \onlinecite{DiCarlo2010}) and charging energy $E_C/h \approx 330 \mhz$ for each qubit.

Details of experimental setup and control are available in a companion publication \cite{DiCarlo2010}.

\subsection{Readout Optimization Procedure}

Setting up the Jaynes-Cummings readout in a cQED device is straightforward.  The bare cavity frequency can be first found by measuring transmission through the device as a function of frequency, using a very large drive power.  It is wise to verify that the measurement amplifier chain will not be saturated by full transmission of this signal ahead of time.  Alternatively, if the $T_C$ of the cavity material is higher than the $T_C$ of the qubit material, the bare cavity frequency can be measured with low power at a temperature between their transition temperatures.  One should take care to account for the effects of kinetic inductance, however, which can be minimized by choosing a temperature as low as possible.  

Once $f_{\mathrm{bare}}$ is ascertained, the optimal readout power for a given qubit detuning can be found in either of two ways.  The first and more straightforward way is to prepare the qubit in either the ground or excited state and measure transmitted amplitude as a function of drive power.  The optimal power can be taken to be that which maximizes the difference between the transmission for the ground or excited qubit states.  In the case of multi-qubit readout, each basis state (or just the ones likely to limit fidelity) should be prepared and measured.  This method has the advantage of being fast and easily automated, though it does not necessarily optimize measurement fidelity per se, but rather measurement contrast.

The second way to find the optimal readout power is to directly measure readout fidelity as a function of power.  At each power, an ensemble of single-shot measurement transients for both prepared states should be integrated, and their fidelities computed as described in the main text (see Fig. 2).  This method is unlikely to give a substantially different optimal drive power than the first method and is substantially more involved, but is technically the correct way of optimizing the conventional definition of measurement fidelity.


\begin{figure*}
\centering
\includegraphics[scale=1]{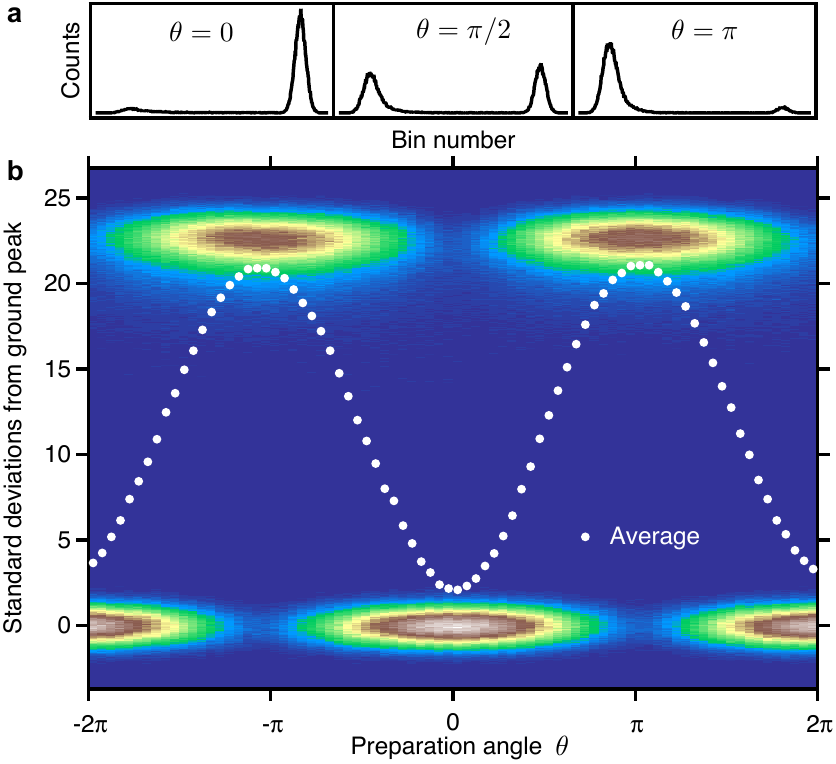}
\caption{
	Quantum measurement following a Rabi oscillation to demonstrate the Jaynes-Cummings readout scheme.
	The qubit is prepared in the superposition state $\bra{\Psi} = \mathrm{cos}(\theta/2)\bra{0} + \mathrm{sin}(\theta/2)\bra{1}$ and repeatedly measured as a function of $\theta$.  In this case, $\theta$ is chosen by scaling the amplitude of a gaussian-shaped microwave pulse at the transition frequency of the qubit. 
	(a) The distribution of integrated values are shown for $\theta=0$, $\theta=\pi/2$, and $\theta=\pi$.  The distribution is always bimodal with very few counts between the two peaks for all $\theta$, corresponding to the system either staying in its dark state or becoming bright almost immediately, demonstrating the single-shot nature of the readout.
	(b) Measurement response distribution for preparation angles between $\theta=-2\pi$ and $\theta=2\pi$.  The $y$-axis is centered on and scaled to the standard deviation ($\sigma$) of the ground state distribution, showing a separation between peaks of about 23$\sigma$ for this integration time of 500~ns.  The average value is plotted (white dots), showing the canonical sinusoidal Rabi oscillation.
	}
{\label{fig:histtheta}}
\end{figure*}
\twocolumngrid

\bibliographystyle{apsrev4-1}
\bibliography{readout_etal}